\documentclass[12pt,a4paper]{article}
\usepackage{amsmath,amsthm,amsfonts}
\newcommand{\ppder}[1]{\frac{\partial^2}{{\partial #1}^2}}
\newcommand{\der}[1]{\frac{d}{d #1 }}
\newcommand{\pder}[1]{\frac{\partial}{\partial #1 }}
\def\req#1{(\ref{#1})}

\date{ }
\author{I. Klich$^{a}$, J. Feinberg$^{a,b}$, A. Mann$^{a}$ and M. Revzen$^{a}$}
\begin{document}
\title{Casimir energy of a dilute dielectric ball with uniform
velocity of light at finite temperature.} \maketitle \vskip 2mm
\begin{center}

$^{a)}$\footnote{e-mail: klich@tx.techion.ac.il; joshua, ady,
revzen@physics.technion.ac.il}Department of Physics, \\ Technion -
Israel Institute of Technology, Haifa 32000 Israel \\
$^{b)}$\footnote{permanent address} Department of Physics, \\
Oranim-University of Haifa, Tivon 36006, Israel\\

\end{center}
\begin{abstract}
The Casimir energy, free energy and Casimir force are evaluated,
at arbitrary finite temperature, for a dilute dielectric ball with
uniform velocity of light inside the ball and in the surrounding
medium. In particular, we investigate the classical limit at high
temperature. The Casimir force found is repulsive, as in previous
calculations.
\end{abstract}
\bigskip


\section{Introduction}

Calculations of the Casimir energy for spherically symmetric
boundary conditions have been the issue of numerous papers, since
the first calculation by Boyer \cite{boyer68} (see also
\cite{balian78,davies72,milton78,nesterenko97,barton99}). Although
a general theory describing the fluctuations of the
electromagnetic field at finite temperature has been present for a
long time
\cite{Lifshitz56,Schwinger70,Schwinger78,Landau,Lifshitz}, the
extension of the results for the dielectric ball from zero
temperature to finite temperatures received less attention. The
case of a conducting spherical boundary (at zero and finite
temperature) was studied by Balian and Duplantier \cite{balian78},
and will serve here to verify the validity of the results we
obtain. Brevik and Clausen \cite{BrevikClausen89}, used the Debye
asymptotic expansion to calculate the Casimir force on the dilute
dielectric ball, assuming uniform velocity of light inside and
outside the ball (including, however, frequency dependence of the
permeability and permittivity), to leading order in the so--called
dilute approximation (the nature of this approximation is
explained following Eq.\req{elH}). It turns out that the usual
Debye approximation is most useful for zero temperature
calculations, but involves complications at finite temperatures.
For example, it was shown \cite{TarekBrevik} that use of these
asymptotics may lead to problematic results, such as that the
Casimir energy of a dielectric ball is independent of temperature.

In this Letter we calculate the thermodynamic features of the
Casimir effect for the dilute dielectric ball at finite
temperature (to leading order in the dilute approximation). We
obtain explicit simple expressions for the Casimir energy, free
energy and for the Casimir force at arbitrary temperature. In
particular, we find that (within the approximation made) the
Casimir force is repulsive for any radius at any temperature (see
Eq.\req{force}). At high temperature its leading behavior is
proportional to ${T \over 16 a^3 \pi}$, where $a$ is the radius of
the ball. Higher order corrections to the dilute approximation
decay like powers of ${1\over T}$, and consequently the leading
behavior ${T \over 16 a^3 \pi}$ gives the Casimir force quite
accurately at high temperature (see section 5).

The starting point of the present calculation is an explicit
expression for the Casimir energy density (to leading order in the
dilute approximation). This expression was derived in \cite{Klich}
using simple properties of the Green's function of the Helmholtz
equation. The Casimir energy at finite temperature, $E_C(T)$, is
then obtained by summing that explicit expression for the energy
over the Matsubara frequencies. Then, using the usual
thermodynamic relation between the Casimir free energy $F_C(T)$
and the Casimir energy $E_C(T)$, we obtain an explicit expression
for $F_C(T)$ valid for all temperatures. We verify that $E_C(T)$
and $F_C(T)$ coincide at $T=0$, as they should (and the Casimir
entropy vanishes at $T=0$). Furthermore, at high temperatures,
$F_C(T)$ tends to a term which is proportional to $-T\log (a T)$
(where $a$ is the radius of the ball). This behavior is consistent
with the results of \cite{balian78} and is also consistent with
the recent general discussion of the high--temperature classical
limit in \cite{Feinberg}.

\section{Mode summation at finite temperatures - general considerations}

The eigenfrequencies of the electromagnetic field, subjected to
some boundary conditions, are solutions of a set of characteristic
equations
\begin{equation}
\Delta_{{\bf c}}^k (\omega)=0 \hspace{1cm}
k=1,2,3...\label{characteristic}
\end{equation}
labeled by an index $k$. (The eigenfrequency spectrum is the union
of solutions to all the equations in \req{characteristic}.) The
set of parameters ${\bf c}$ describes the constraints, such as the
radius of the sphere or the distance between the conducting
plates. The functions $\Delta_{{\bf c}}^k (\omega)$ are usually
expressions involving solutions of the scalar Helmholtz equation
subjected to appropriate boundary conditions. In the following we
consider the case of spherical symmetry, where the functions
$\Delta_{{\bf c}}^k (\omega)$ involve combinations of Bessel
functions (see Eq. (\ref{Fldef}-\ref{elH}) in sec. 3).

\begin{figure}[t]\center{\input epsf \epsfbox{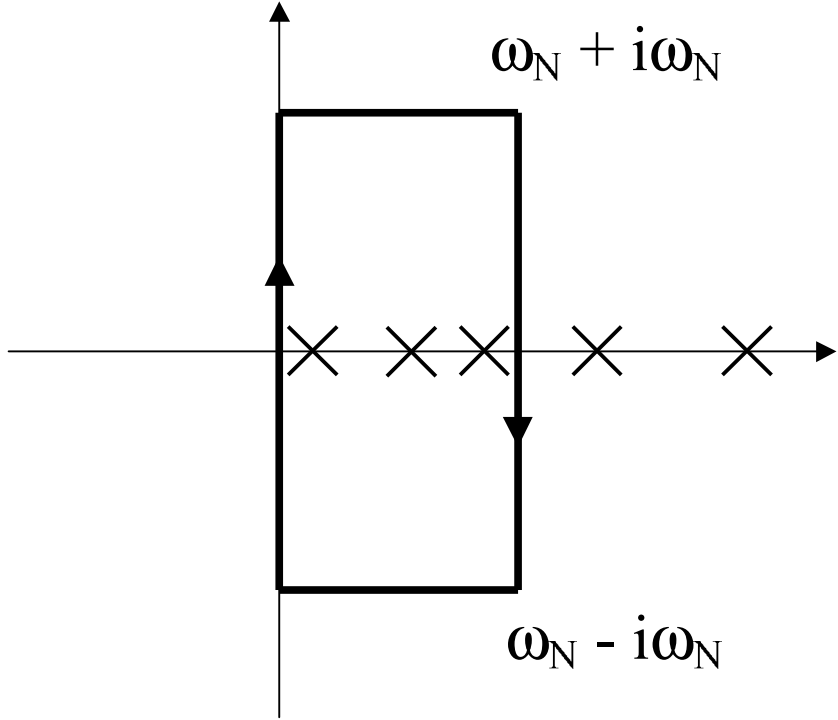}} \caption{A rectangular
contour in the complex plane.} \label{f1}
\end{figure}

Let us first consider the system in its ground state at zero
temperature. The zero temperature Casimir energy is of course the
difference between the formally divergent sum over all allowed
modes of the constrained system and the divergent sum over the
modes of the free electromagnetic field,
\begin{equation}
E_C=\sum_n {\omega_n({\bf c})\over 2}-\sum_n {\omega_n^{\it free
}\over 2}.
\end{equation}
This expression is still divergent and needs regularization. Thus,
we introduce a UV--frequency cut-off $\omega_N$ and define
\begin{equation}
E_C(\omega_N)=\sum_{\omega_n\leq\omega_N} {\omega_n({\bf c})\over
2}-\sum_{\omega_n\leq\omega_N} {\omega_n^{\it free }\over 2}.
\label{defsumcut}
\end{equation}
When the boundaries are pushed to infinity, the system ceases to
be constrained and becomes free. This is described by letting the
parameters {\bf c} in \req{defsumcut} tend to an appropriate
value, say, infinity. It is convenient to represent the sum in the
last expression as a contour integral
\begin{equation}\label{definitionofCas}
E_C(\omega_N)=\sum_k\oint_{\Gamma_N}
\frac{z}{2}\,F_{k}(z;\bold{c})dz
\end{equation}
where $\Gamma_N$ is shown on Fig.\ref{f1}, and
\begin{equation}\label{Dofstates}
F_{k}(z;\bold{c})=\lim_{\bold{c'}\rightarrow {\it
free}}\frac{1}{2\pi i
}\der{z}(\log\Delta_{\bold{c}}^k(z)-\log\Delta_{\bold{c'}}^k (z)),
\end{equation}
is the (regulated) resolvent corresponding to index $k$ in
\req{characteristic}, and ``${\it free}$" denotes the set of
parameters describing the limiting case where the boundaries are
taken to infinity. In the case of spherical geometry this is
achieved by taking the radius to infinity. The Casimir energy is
then defined as
\begin{equation}
E_C=\lim_{\omega_N\rightarrow\infty}E_C(\omega_N).
\end{equation}
To carry over these considerations to finite temperature, we
recall that the average energy per mode of the electromagnetic
field is ${\omega\over 2}\coth(\beta\omega/2)$ and that the free
energy per mode is $\frac{1}{\beta}\log(2\sinh(\beta \omega/2))$,
where $\beta=\frac{1}{k_B T}$. Thus,
\begin{equation}
E_C(T;\omega_{N})=\sum_k \oint_{\Gamma_N}
\frac{z}{2}\,\coth(\frac{\beta z}{2})F_{k}(z;\bold{c})
dz,\label{Econ}
\end{equation}
is the (cut off) Casimir energy at temperature $T$, and
\begin{equation}
F_C(T;\omega_{N})=\sum_k \oint_{\Gamma_N}
\frac{1}{\beta}\log(2\sinh(\beta z/2))F_{k}(z;\bold{c})dz,
\label{free}
\end{equation}
is the (cut off) Casimir free energy at temperature $T$. Upon
rotating the integration contour in \req{Econ} to wrap around the
imaginary axis, the Casimir energy can be expressed as a sum over
Matsubara modes, namely
\begin{equation}
E_C=\sum_{n=-\infty}^{\infty}\sum_k
\left(\frac{z}{2}\,F_{k}(z;\bold{c})\right)\Big|_{z={2\pi i
n\over\beta}} \label{tempsum}.
\end{equation}
This concludes our general considerations.
\section{Casimir energy}

In this section we study the temperature dependence of the Casimir
energy of a dielectric ball, having permittivity and permeability
$\epsilon,\mu$ which is embedded in a medium with permittivity and
permeability $\epsilon',\mu'$. A further simplifying assumption
which we make throughout this Letter, is that the velocity of
light be continuous across the boundary, namely
\begin{equation}\label{equalvelocity}
\epsilon\mu=\epsilon'\mu'.
\end{equation}
This condition is often desirable as it causes the frequency
equations to simplify and some divergences to cancel out (see
\cite{BKBN} and references therein). It turns out that in this
case the energy depends on the dielectric data only through the
parameter $\xi^2$ defined by
\begin{equation}
\xi^2=\left( \varepsilon_1-\varepsilon_2\over
\varepsilon_1+\varepsilon_2 \right)^2 \label{defxi2}
\end{equation}
which is symmetric under
$\varepsilon_1\leftrightarrow\varepsilon_2$. Thus, the Casimir
energy is invariant under interchanging the inner and outer media.

It has been shown \cite{nesterenko97,Klich,BKBN} that under the
condition \req{equalvelocity}, and after rotating the integration
contour in \req{definitionofCas} to the imaginary axis, that the
difference in zero point energy of two balls of radii $a$ and $b$
may be written as
\begin{equation}
E_C(a)-E_C(b)=\sum_{l=1}^{\infty} (l+{1\over
2})\int^{\infty}_{\infty} d\omega \omega F_{l}(\omega;a,b)
\label{energydif},
\end{equation}
where the resolvent \req{Dofstates} is
\begin{equation}\label{Fldef}
F_{l}(\omega;a,b)=-{1\over 2\pi}{d\over
d\omega}\biggr(\log\big[1-\xi^2\lambda_{l}^2(a
|\omega|)\big]-\log\big[1-\xi^2\lambda_{l}^2(b
|\omega|)\big]\biggr).
\end{equation}
Here
\begin{equation}
\lambda_{l}(a|\omega|)=(s_{l}e_{l}(a |\omega|))',
\end{equation}
where $s_{l},e_{l}$ are the modified Bessel
functions ($\nu=l+{1\over 2}$)
\begin{equation}
s_{l}(x)=ixj_l(ix)=\sqrt{\frac{\pi
x}{2}}e^{-\frac{i\pi\nu}{2}}J_{\nu}(ix)
\end{equation}
\begin{equation}\label{elH}
e_{l}(x)=ixh_l^{(1)}(ix)=i\sqrt{\frac{\pi
x}{2}}e^{\frac{i\pi\nu}{2}} H^1_{\nu}(ix)
\end{equation}
The case of $\xi$ finite but small corresponds to the inner ball
being nearly identical in its properties to the surrounding
medium. This is referred to in the literature as the limit of
dilute media. In this limit we can expand \req{Fldef} in powers of
$\xi^2$. This expansion is further justified by the fact that on
the imaginary $\omega$ axis $\lambda_l^2$ is small, thus the
expansion can be pushed to $\xi^2\sim 1$ (for $\xi^2=1$, of
course, one of the media has infinite conductivity and is thus a
perfect conductor). In this Letter we will content ourselves with
the leading term in this expansion. Thus, to order $\xi^2$ we have
from \req{Fldef}
\begin{equation}
F_{l}(\omega;a,b)={\xi^2\over 2\pi}{d\over
d\omega}\biggr(\lambda_{l}^2(a |\omega|)-\lambda_{l}^2(b
|\omega|)\biggr)\label{Fldefexp},
\end{equation}

In \cite{Klich} one of us was able to carry the sum over $l$ in
\req{energydif} to first order in $\xi^2$ in closed form, without
using Debye approximations, as is usually done in the literature.
The result was
\begin{equation}
\xi^2 \sum_{l=1}^\infty (l+\frac{1}{2})x\der{x}\lambda_l^2(x)=
\frac{\xi^2}{2}(-\frac{1}{2}+\frac{1}{2}e^{-4x}(1+2x)^2).
\label{sumI}
\end{equation}
Thus, the relative resolvent, to order $\xi^2$, is
\begin{equation}
F^{(2)}(z;a,b)=\frac{\xi^2}{4 \pi z}\left[e^{-4a|z|} (1+4a|z|+4a^2
|z|^2)
-e^{-4b|z|}(1+4b|z|+4b^2|z|^2)\right]\label{relativeDensity}
\end{equation}
(here $z$ is a pure imaginary frequency.)

Using this simple explicit expression, we can calculate the energy
in the dilute limit at an arbitrary temperature. In order to
calculate the zero point energy with respect to the vacuum we take
the limit $b\rightarrow \infty$ at the outset. Note that the $x$
independent term in \req{sumI} cancels between the two terms in
the difference \req{relativeDensity}, resulting in an expression
which decays exponentially when $|z|\rightarrow\infty$. Thus, no
further regularization will be needed when we integrate over
$\omega$. Using \req{tempsum} and \req{relativeDensity}, the
Casimir energy at temperature $T$ is
\begin{eqnarray}
& E_C(a,T)=\frac{\xi^2 T}{4}\sum_{n=-\infty}^{\infty}e^{-8 a T \pi
|n|} (1+8 a T \pi |n|+16 a^2 T^2 \pi^2 n^2)=\\ \nonumber &
\frac{\xi^2 T}{4}\left(1-T\pder{T}+{T^2\over
4}\ppder{T}\right){2\over e^{8a \pi T}-1}+\frac{\xi^2 T}{4},
\end{eqnarray}
yielding
\begin{eqnarray}\label{ballenergy}
& E_C(a,T)=\frac{T\xi^2}{4}+\frac{T\xi^2}{2\left( e^{8a\pi
T}-1\right)}+\frac{4a\pi T^2\xi^2e^{8a\pi T}} {{{\left(  e^{8a\pi
T}-1\right) }^2} }+ \\ \nonumber &
\frac{8a^2\pi^2{T^3}\xi^2e^{8a\pi T}\left(1+e^{8a\pi T}
\right)}{{\left(  e^{8a\pi T}-1\right) }^3}
\end{eqnarray}
This result is exact (to order $\xi^2$) for all temperatures. It
is immediate to check that as $T$ goes to zero we retain the
correct zero temperature result \cite{Klich}
\begin{equation}
E_C(a,0)=\frac{5\,\xi^2}{32a\pi }\label{zero}\,\,\,.
\end{equation}
Note that at high temperatures, $T>>{1\over 8\pi a}$, we have
\begin{equation}
E_C(a,T)=\frac{T\xi^2}{4}+{\cal O}( e^{-8a\pi T})\label{high},
\end{equation}
which is consistent with Eq.(8.39) in \cite{balian78} for the free
energy, to leading order. Namely, the Casimir energy at high
temperatures is independent of the radius of the ball. That
$E_C(T,a)$ becomes independent of the radius at high temperature
is expected on general grounds (see \cite{Feinberg} for a detailed
discussion): At high enough temperatures we expect classical
equipartition to hold. Since there is obviously a one to one
correspondence between the states of the system with radius $a$
and the states of any other system with a different radius $b$,
the difference between the zero point energies of these two
systems must vanish at high temperatures. Indeed, for the case
studied here, the difference (to order $\xi^2$) vanishes
exponentially fast with temperature.

\section{Free energy and force}

Having calculated the energy we may now use the expression
\req{ballenergy} to infer the Casimir free energy of the system.
From the general relation
\begin{equation}
E={Tr(He^{-\beta H})\over Tr(e^{-\beta H})} =-\pder{\beta}(\log
Z)=\pder{\beta}(\beta F ) \label{EfromF},
\end{equation}
we have
$ F={1\over\beta}\int E(\beta) d \beta+{C\over\beta}$
(where $C$ is an integration constant to be determined), or,
equivalently,
\begin{equation}
F=-T\int {E(T)\over T^2} dT+C T\,\,. \label{FfromE}
\end{equation}
To evaluate the Casimir free enrgy at low temperatures we first
expand $E_C(T)$ around $T=0$
\begin{equation}
E_C(a,T)=\frac{5\,\xi^2}{32a\pi
}+\frac{8a^3\pi^3\xi^2{T^4}}{45}+\frac{512a^5\pi^5\,\xi^2\,{T^6}}{945}
+{\cal O}(T^7) \label{seriesE}
\end{equation}
(Note that the first temperature dependent contribution to the
energy is of order $T^4$.) Thus, from \req{FfromE}, the Casimir
free energy is $$F_C(a,T)=CT
+\frac{5\,\xi^2}{32a\pi}-\frac{8a^3\pi^3T^4\xi^2}{135} -
\frac{512a^5\pi^5T^6\xi^2}{4725}+{\cal O}(T^7)\,.$$ To fix the
constant $C$, we use the third law of thermodynamics, namely, that
the Casimir entropy
\begin{equation}\label{entropy}
S_C(a,T)={E_C(a,T)-F_C(a,T)\over T}
\end{equation}
vanishes at $T=0$. It is straightforward to check that this
implies $C=0$. Therefore, at low temperatures, the Casimir free
energy is:
\begin{equation}
F_C(a,T)=\frac{5\,\xi^2}{32a\pi }-\frac{8a^3\pi^3T^4\xi^2}{135}
-\frac{512a^5\pi^5T^6\xi^2}{4725}+O(T^7)\,\,.
\end{equation}
This result coincides with the two scattering approximation
\cite{balian78} for the conducting sphere (up to multiplication by
$\xi^2$).


\noindent To proceed beyond low temperatures we rewrite
\req{FfromE} as
\begin{equation}
F=-T\int_0^T {E(t)-E(0)\over t^2} dt+E(0) \label{EtoF}
\end{equation}
One can easily check, using the power series \req{seriesE}, that
the integral on the right side converges, and that indeed $F$
satisfies the condition \req{EfromF}. To show that there is no
additional term of the form $CT$ in \req{EtoF}, we note that this
expression can also be verified for any particular mode $\omega$
of the electromagnetic field. Writing,
\begin{equation}
T\log(2\sinh{\omega\over 2T})=-T\int_0^T {{\omega\over
2}\coth{\omega\over 2t}-{\omega\over 2}\over t^2} dt+{\omega\over
2} \label{EtoFperMode},
\end{equation}
and then summing over the modes gives precisely \req{EtoF}.

In order to evaluate \req{EtoF} it is convenient to split $E_C$
into
\begin{equation}
E_C(a,T)=E_{C1}(T)+E_{C2}(T)
\end{equation}
where
\begin{equation}
E_{C1}(T)=\frac{4a\pi T^2\xi^2e^{8a\pi T}} {{{\left(e^{8a\pi
T}-1\right) }^2} }+ \frac{8a^2\pi^2{T^3}\xi^2e^{8a\pi
T}\left(1+e^{8a\pi T} \right)}{{\left(e^{8a\pi T} -1
  \right) }^3} \label{E1}
\end{equation}
and
\begin{equation}
E_{C2}(T)=\frac{T\xi^2}{4}+\frac{T\xi^2}{2\left( e^{8a\pi
T}-1\right)} \label{E2}
\end{equation}
and then split the integral in \req{EtoF} accordingly. The
contribution associated with $E_{C1}$ can be integrated in a
straightforward manner, whereas the contribution from $E_{C2}$
requires more careful treatment. Thus,
\begin{equation}\label{FreeFromE}
F_C=-T\int_0^T (E_{C1}(t)-E_{C1}(0)){dt\over t^2} -T\int_0^T
(E_{C2}(t)-E_{C2}(0)){dt\over t^2} +\frac{5\,\xi^2}{32a\pi }
\end{equation}
The term associated with $E_{C1}$ is now straightforwardly
integrated to yield
\begin{equation}
F_1(T)=\xi ^2\left[-{\frac{3}{32\,a\,\pi }}+{\frac{5\,{T
}}{16}}+{\frac{a\,\pi \,{T^2}}
   {{{\left({e^{8\,a\,\pi \,T}} -1  \right) }^2}}}+{
    \frac{\left( 5\,T\,+ 8\,a\,\pi \,{T^2}\,
       \right)}{8\,\left({e^{8\,a\,\pi \,T}}-1 \right) }}\right]. \label{E1T}
\end{equation}
We are now left with the second term, which can be written as
\begin{equation}
F_2(T)=-T\int_0^T (E_{C2}(t)-\frac{3\,\xi^2}{32a\pi }){dt\over
t^2} =-{T\xi^2\over 2}\int_0^{8 a \pi T}\left({1\over 2}-{1\over
t}+{1\over e^t-1}\right){dt\over t}
\end{equation}
The last integral is, of course, well defined at $T=0$, but
diverges logarithmically as $T\rightarrow\infty$. To simplify this
expression we observe that
\begin{equation}
\int_0^{\infty}\left({1\over 2}-{1\over t}+{1\over
e^t-1}\right){e^{-z t}\over t}dt=\log\Gamma(z)-z\log z+z+{1\over
2}\log z-{1\over 2}\log 2\pi
\end{equation}
(eq. 8.341.1 in \cite{ryzhik}), where $z>0$ can be thought of as a
UV--regulator, which is set to zero in the end (it is understood
that $8 a \pi T z<<1$ throughout the calculations.). Thus, writing
\begin{eqnarray}
&& \int_{0}^{8 a \pi T}\left({1\over 2}-{1\over t}+{1\over
e^t-1}\right){1\over t}dt= \\ \nonumber && \lim_{z\rightarrow
0}\int_0^{8 a \pi T}\left({1\over 2}-{1\over t}+{1\over
e^t-1}\right){e^{-z t} \over t}dt
\end{eqnarray}
and using the expansions
\begin{equation}
\log\Gamma(z)=-\log z-\gamma z+{\cal O}(z^2)
\end{equation}
(6.1.33 in \cite{abramowitz64}, $\gamma=0.577...$ is Euler's
constant) and
\begin{equation}
\int_{8 a \pi T}^{\infty}{e^{-z t}\over t}dt=-\gamma-\log (8 a \pi
T z )+{\cal O}(8 a \pi T z),
\end{equation}
(eq. 8.214.1 in \cite{ryzhik}), we find after taking the limit
$z\rightarrow 0^{+}$, that
\begin{eqnarray}
& \int_{0}^{8 a \pi T}\left({1\over 2}-{1\over t}+{1\over
e^t-1}\right){1\over t}dt=  \\ \nonumber \\ \nonumber & {1\over
2}\log (4 a T )+{\gamma\over 2}+{1\over 8 a \pi T}-\int_{8 a \pi
T}^{\infty}{1\over (e^t-1)t}dt\, ,
\end{eqnarray}
in which the last integral is well defined for all $T$. Combining
this result with \req{E1T}, we obtain from \req{FreeFromE} the
following expression for the Casimir free energy at arbitrary
temperature:
\begin{eqnarray}\label{freeFinal}
& F_C(a,T)=-{T\xi^2\over 4}\left[\log (a T )+\log
4+\gamma-{5\over4}\right]+\\ \nonumber &{T\xi^2\over 2}\int_{8 a
\pi T}^{\infty}{dt\over (e^t-1)t}+ {\frac{a\,\pi \,{T^2}\,\xi ^2}
   {{{\left({e^{8\,a\,\pi \,T}} -1  \right) }^2}}}+{
    \frac{\left( 5\,T\,\xi ^2 + 8\,a\,\pi \,{T^2}\,\xi ^2
       \right) }{8\,\left({e^{8\,a\,\pi \,T}}-1 \right) }}
\end{eqnarray}
Note that $\log 4+\gamma-{5\over4}=0.714$, and thus, the linear
and $T\log (a T)$ terms in \req{freeFinal} agree with the
two-scattering approximation \cite{balian78} (when $\xi=1$). All
other terms in \req{freeFinal} decay exponentially as
$T\rightarrow\infty$ and not as powers as is seen in eq. 8.39
\cite{balian78}. This is due to the fact that the density of
states in the $\xi^2$ approximation decays exponentially with
frequency and not as $1/\omega^2$, which was the basis of the high
temperature approximation used in \cite{balian78} (see eq. (6.12)
therein) . This suggests that at sufficiently high temperatures,
we will have to include higher order corrections in $\xi^2$ to see
the power law decay. However, as we shall see in the next section,
there are no further corrections to the $T$ and $T\log (a T)$
terms which depend on the radius $a$. Thus, the $\xi^2$
approximation gives a reliable estimate of the Casimir force at
high temperatures. Another interesting feature of the last
expression is that $F_C(T)<0$ for $a$ large, while $F_C(T)>0$ for
small values of $a$. Thus, for any temperature $T$ there is an
intermediate radius $b(T)$ such that $F_C=0$ for this radius (it
follows that the free energy may be also viewed as relative to the
energy of the ball with this zero-energy radius.)

From \req{freeFinal} we calculate the Casimir force to order
$\xi^2$
\begin{eqnarray}\label{force}
&& {\cal{F}}_C(a,T)=-{1\over 4\pi a^2}\pder{a}F_C(a,T)=\\
\nonumber && {T \xi^2\over 16 a^3 \pi}+{T \xi^2\over 8 a^3 \pi
(e^{8 a \pi T}-1)}+{ T^2 \xi^2 e^{8 a \pi T} \over a^2 (e^{8 a \pi
T}-1)^2}+{2 \pi T^3 \xi^2(e^{8 a \pi T}+1) e^{8 a \pi T} \over a
(e^{8 a \pi T}-1)^3} \,\,.
\end{eqnarray}
We see that the force \req{force} is positive for all values of
$a$ and $T$. Moreover it behaves as ${T \xi^2\over 16 a^3 \pi}$ at
high temperatures. Another quantity of interest, which is the
Casimir entropy \req{entropy}, behaves as ${\xi^2\over
4}(1.714+\log (a T))$ at high temperatures.


\section{Relation to direct mode summation}

In this section we use a different approach based on evaluating
\req{free} directly. This analysis was carried out for high and
low temperatures in the case of the conducting sphere by Balian
and Duplantier \cite{balian78}. To make our discussion well
defined we consider the difference in free energy between two
balls of radii $a$ and $b$. From \req{free} we deduce that the
relative Casimir free energy is
\begin{equation}
F_C(a)-F_C(b)=\sum_{l=1}^{\infty} (l+{1\over 2})\frac{i}{\beta
}\int^{\infty}_{-\infty}F_{l}(\omega;a,b)\log(2i\sin(\frac{\beta
\omega}{2})) d\omega. \label{Freenergydif}
\end{equation}
It is convenient to decompose the logarithm as
\begin{eqnarray}\label{log}
& \log\left(2i\sin (\frac{\beta \omega}{2})\right)= \\ \nonumber &
\log\left(2\left|\sin(\frac{\beta
\omega}{2})\right|\right)+i\pi\sum_{n = 0}^{'\infty} \theta
\left(\omega-\frac{2\pi n}{\beta }\right)-i\pi\sum_{n =
0}^{'\infty}\theta\left(-\omega-\frac{2\pi n}{\beta }\right)
\end{eqnarray}
(where, as usual, $\sum_{n=0}^{'\infty}$ means that the $n=0$ term
is counted with half-weight). Note that $F_{l}(\omega;a,b)$ is an
odd function of frequencies. Thus, discarding the even part of
\req{log}, we are left with
\begin{eqnarray}\label{expforfree}
& F_C(a)-F_C(b)= \frac{2\pi}{\beta }\sum_{l=1}^{\infty} (l+{1\over
2})\sum_{n=1}^{\infty}\int_{{2\pi \,n\over\beta }}^{\infty}
F_{l}(\omega;a,b)d\omega+\\ \nonumber &  \frac{\pi}{\beta
}\sum_{l=1}^{\infty} (l+{1\over 2})\int_{0}^{\infty}
F_{l}(\omega;a,b)d\omega\,.
\end{eqnarray}
The first term on the right side is a decaying function of
temperature (it decays as ${1\over T}$). The linear and
logarithmic contributions in $T$ come only from the second term.

We now expand \req{expforfree} in powers of $\xi^2$, and work to
leading order in $\xi^2$. Substituting $F^{(2)}(\omega;a,b)$ from
\req{relativeDensity} into \req{expforfree}, we can show that the
leading term in \req{expforfree} is consistent with the expression
\req{freeFinal} we obtained in the previous section. (In the
following we integrate without using the $b$ dependent terms,
wherever they are not necessary for convergence.) We start with
the easily integrable expressions:
\begin{eqnarray}
&
\frac{\xi^2}{2\beta}\sum_{n=1}^{\infty}\int_{\frac{2n\pi}{\beta}}^{\infty}
(4a e^{-4a\omega}+ 4a^2 \omega e^{-4a\omega})d\omega
=\frac{\xi^2}{2\beta}\sum_{n=1}^{\infty}\frac{e^{\frac{8an\pi}
{\beta}}(8an\pi+5\beta)} {4\beta} \\ \nonumber &
=
\frac{\xi^2}{8}\frac{8a\pi
e^{\frac{8a\pi}{\beta}}+5\beta(e^{\frac{8a\pi}{\beta}}-1)}{
\beta^2(-1+e^{\frac{8a\pi}{\beta}})^2}
\end{eqnarray}
Note that we summed starting with $n=1$, since the $n=0$ term is
independent of the radius and thus cancels out (when taking the
difference). The less trivial terms can be calculated as follows
\begin{eqnarray}
&&-\frac{\xi^2}{2\beta }\sum_{n=0}^{\infty}{'}\int_{\frac{2 n\pi
}{\beta }}^{\infty
}\frac{1}{\omega}\left(e^{-4a\omega}-e^{-4b\omega}\right)d\omega
\\ \nonumber &&=-\frac{\xi^2}{2\beta
}\sum_{n=0}^{\infty}{'}\int_{\frac{2 n\pi }{\beta }}^{\infty
}\int_{a}^{b}4e^{-4u\omega}du d\omega=-\frac{\xi^2}{2\beta
}\int_{a}^{b}\sum_{n=0}^{\infty}{'} \frac{1}{u}e^{-\frac{8 u n\pi
}{\beta }}du =\\ \nonumber &&
-\frac{\xi^2}{2\beta}\left(-\frac{1}{2}\log({a\over
b})+\int_{\frac{8a\pi }{\beta
}}^{\infty}\frac{1}{u(-1+e^{u})}du-\int_{\frac{8b\pi }{\beta
}}^{\infty}\frac{1}{u(-1+e^{u})}du\right)\,\,.
\end{eqnarray}

Thus the difference between the free energy of two balls is:
\begin{equation}
F_C(a,T)-F_C(b,T)=-\frac{T \xi^2}{2}\left[\frac{1}{2}\log({a\over
b})+M(8\pi a T)-M(8\pi b T)\right] \label{freeball}
\end{equation}
where
\begin{equation}
M(x)=\frac{xe^{x}+5(-1+e^{x})}{4{\left(-1+e^{x}\right)}^2}+
\int_{x}^{\infty}\frac{1}{u(-1+e^{u})}du \label{lfun}
\end{equation}
Clearly \req{freeball} and \req{freeFinal} are consistent.

Finally, we make the important observation that terms of order
$\xi^4$ and higher in the expansion of \req{expforfree} do not
contribute to $F_C(a,T)$ terms that are proportional to $T$ or
$T\log (a T)$ with coefficients that depend on $a$. To see this,
note from \req{Fldef} and from \req{expforfree} that contribution
to terms proportional to $T$ or $T\log (a T)$ will come from
expression such as
\begin{eqnarray}
& \frac{\pi}{\beta }\sum_{l=1}^{\infty} (l+{1\over
2})\int_{0}^{\infty} F^{(2m)}_{l}(\omega;a,b)d\omega=\\ \nonumber
& \frac{\pi}{\beta }\sum_{l=1}^{\infty} (l+{1\over
2})\int_{0}^{\infty} {\xi^{2m}\over 2\pi m}{d\over
d\omega}\biggr(\lambda_{l}^{2m}(a |\omega|)-\lambda_{l}^{2m}(b
|\omega|)\biggr)
\end{eqnarray}
(The other terms, namely, terms coming from the first sum in
\req{expforfree}, obviously decay as powers in ${1\over T}$.) For
any $m>1$ each term in the last sum can be integrated to yield:
\begin{equation}
\lim_{\omega\rightarrow 0^+}\frac{\pi}{\beta }\sum_{l=1}^{\infty}
(l+{1\over 2}){\xi^{2m}\over 2\pi m}\biggr(\lambda_{l}^{2m}(a
\omega)-\lambda_{l}^{2m}(b
\omega)\biggr)=0
\end{equation}
since for all $m>1$ this sum is convergent at $\omega=0$ (note
that $\lambda_l(0)={1\over 2l+1}$). Thus, the $T$ and $T\log (a
T)$ terms which depend on $a$ in the {\it exact} $F_C(a,T)$ are
accurately accounted for by the order $\xi^2$ expression
\req{freeFinal}. Consequently, \req{force} gives the Casimir force
quite accurately at high $T$, up to corrections in ${1\over T}$.

\section{Conclusions}

The temperature dependence of the Casimir energy, free energy and
force were studied in the dilute dielectric ball approximation. We
showed that the free energy, and thus the force, can be obtained
from the Casimir energy, which is an easier quantity to calculate.
This was done for an arbitrary temperature, and agrees with the
two-scattering results for a conducting shell obtained in
\cite{balian78} for the high and low temperature limits. We
confirmed that in the high temperature limit the free energy
behaves as $-T\log (a T)$, while the Casimir energy is linear in
$T$. However this linear term is independent of geometry, and
thus, the energy difference between two balls of different radii
decays exponentially. This is expected from the equipartition
property of classical statistical mechanics - at high temperatures
the average energy per mode, $k_B T$, is independent of the
frequency, and since there is a one to one correspondence between
the eigenmodes of two spheres of different radii, the difference
in the energy should vanish in the high temperature limit. The
calculations were based on the method developed in \cite{Klich}.
This method have since been used in other related problems at zero
temperature in \cite{Klcyl,Ma,Lambiase}. We expect that these
cases now too admit an exact treatment, along similar lines as
presented here.


\section*{Acknowledgments}
I. K. and M. R. thank I. Brevik for getting them interested in
this problem. M. R. also thanks I. Brevik for his kind hospitality
at Trondheim during summer 1998. I. K. wishes to thank A. Elgart
for discussions. This work was supported by the Technion-Haifa
University Joint Research Fund. The work of A. M. and M. R. was
supported in part by the Technion VPR Fund, and by the Fund for
Promotion of Research at the Technion. J. F.'s research was
supported in part by the Israeli Science Foundation grant number
307/98(090-903).
 

\noindent

\begin{thebibliography}{99}

\bibitem{boyer68}
T. H. Boyer, Phys. Rev. {\bf 174}, 1764 (1968).

\bibitem{balian78}
R. Balian and B. Duplantier, Ann. Phys. (N.Y.) {\bf 112}, 165
(1978).

\bibitem{davies72}
B. Davies, J. Math. Phys. {\bf 13}, 1324 (1972).

\bibitem{milton78}
K. A. Milton, L. L. DeRaad, Jr., and J. Schwinger, Ann. Phys.
(N.Y.) {\bf 115}, 388 (1978).

\bibitem{nesterenko97}
V. V. Nesterenko and I. G. Pirozhenko, Phys. Rev. D {\bf 57}, 1284
(1997).

\bibitem{barton99}
G. Barton, J. of Phys. A {\bf 32}, 525 (1999).

\bibitem{Lifshitz56}
E. M. Lifshitz, Sov. Phys. JETP {\bf 2} 73, (1956).

\bibitem{Schwinger70}
J. Schwinger, {\it Particles, Sources and Fields} (Addison-Wesley,
Reading, MA 1970) Vol 1.

\bibitem{Schwinger78}
J. Schwinger, L. L. DeRaad, Jr. and K. A. Milton, Ann. Phys.
(N.Y.) {\bf 115}, 1 (1978).

\bibitem{Landau}
L. D. Landau, E. M. Lifshitz and L. P. Pitaevskii, {\it
Electrodynamics of continuous media} 2nd ed. (Pergamon, Oxford,
1984)

\bibitem{Lifshitz}
E. M. Lifshitz and L. P. Pitaevskii, {\it Statistical Mechanics},
Part 2 (Pergamon, Oxford, 1984)

\bibitem{BrevikClausen89}
I. Brevik and I. Clausen, 
Phys Rev D, {\bf 39}(2):603-611 (1989)

\bibitem{TarekBrevik}
T. Ali Yousef and I. Brevik unpublished.

\bibitem{Klich}
I. Klich, {\it Casimir's energy of a conducting sphere and of a
dilute dielectric ball}, Phys. Rev. D {\bf 61} (2000) 025004,
hep-th/9908101.

\bibitem{Feinberg}
J. Feinberg, A. Mann, M. Revzen.{\it Casimir Effect: The Classical
Limit} hep-th/9908149.

\bibitem{BKBN}
I. Brevik and H. Kolbenstvedt, {\it Phys. Rev. }{\bf D25} (1982)
1731; {\it Ann. Phys. } (NY) {\bf 143} (1982) 179; {\bf 149}
(1983) 237; I. Brevik and G.H. Nyland, {\it Ann. Phys. } (NY) {\bf
230} (1984) 321.

\bibitem{Klcyl}
I. Klich and A. Romeo {\it Infinite dielectric cylinder subject to
equal light-velocity constraint}, (submitted for publication in
Phys Lett. B.)

\bibitem{Ma}
V. Marachevsky, {\it Casimir-Polder energy and dilute dielectric
ball: nondispersive case}, hep-th/9909210.

\bibitem{Lambiase}
G. Lambiase, G. Scarpetta and V. V. Nesterenko, {\it Exact value
of the vacuum electromagnetic energy of a dilute dielectric ball
in the mode summation method}, hep-th/9912176.

\bibitem{ryzhik}
I. S. Gradshtein and I. M. Ryzhik, {\it Table of integrals, series
and products}, 5th edition (Academic Press, N.Y. 1994.)

\bibitem{abramowitz64}
 {\it Handbook of Mathematical Functions}, (Natl. Bur. Stand. Appl. Math. Ser.
55), edited by M. Abramowitz and I. A. Stegun (U. S. GPO,
Washington, D.C., 1964) ( reprinted by Dover, New York, 1972).

\end{thebibliography}
\end{document}